\documentclass[journal=jacsat,manuscript=article]{achemso}
\usepackage[version=3]{mhchem} 
\usepackage{siunitx}
\usepackage[dvipsnames]{xcolor}
\usepackage{subcaption}

\author{Carolina Paba}
\affiliation{Department of Physics, University of Trieste, 34127 Trieste, Italy}
\author{Virginia Dorigo}
\affiliation{Hochschule Fresenius, 65510 Idstein, Germany}
\author{Beatrice Senigagliesi}
\affiliation{Elettra Sincrotrone Trieste, 34149 Basovizza TS, Italy}
\alsoaffiliation{Current address: IINS, Bordeaux Neurocampus, 33076 Bordeaux Cedex, France}
\author{Nicolò Tormena}
\affiliation{Department of Physics, University of Durham, Durham DH1 3LE, United Kingdom}
\author{Pietro Parisse}
\affiliation{IOM-CNR, 34149 Basovizza TS, Italy}
\alsoaffiliation{Elettra Sincrotrone Trieste, 34149 Basovizza TS, Italy}
\email{parisse@iom.cnr.it}
\phone{+39 3756416}
\author{Kislon Voitchovsky}
\email{kislon.voitchovsky@durham.ac.uk}
\affiliation{Department of Physics, University of Durham, Durham DH1 3LE, United Kingdom}
\phone{+44 191 334 3615}
\author{Loredana Casalis}
\email{loredana.casalis@elettra.eu}
\affiliation{Elettra Sincrotrone Trieste, 34149 Basovizza TS, Italy}
\phone{+39 040 375 8291}
\title[An \textsf{achemso} demo]
  {Lipid bilayer fluidity and degree of order regulates small EVs adsorption on model cell membrane}
\abbreviations{AFM,sEVs,SLB,$L_d$,$L_o$,$S_o$,Chol,DOPC,SM,UC-MSC,MDA-MB-231,TNBC}

\begin{document}

\begin{abstract}
Small extracellular vesicles (sEVs) are known to play an important role in the communication between distant cells and to deliver biological information throughout the body. To date, many studies have focused on the role of sEVs characteristics such as cell origin, surface composition, and molecular cargo on the resulting uptake by the recipient cell. Yet, a full understanding of the sEV fusion process with recipient cells and in particular the role of cell membrane physical properties on the uptake are still lacking. Here we explore this problem using sEVs from a cellular model of triple-negative breast cancer fusing to a range of synthetic planar lipid bilayers both with and without cholesterol, and designed to mimic the formation of ‘raft’-like nanodomains in cell membranes. Using time-resolved Atomic Force Microscopy we were able to track the sEVs interaction with the different model membranes, showing the process to be strongly dependent on the local membrane fluidity. The strongest interaction and fusion is observed over the less fluid regions, with sEVs even able to disrupt ordered domains at sufficiently high cholesterol concentration. Our findings suggest the biophysical characteristics of recipient cell membranes to be crucial for sEVs uptake regulation.
\end{abstract}
\section{Introduction}

%
The plasma membrane has an essential role in maintaining cell homeostasis\cite{tekpli2013role}, and can actively  regulate its molecular composition and shape in response to external stimuli\cite{sezgin2017mystery, simons2011membrane}.
In particular, it is known that some dynamic membrane microdomains such as lipid rafts and caveolae contribute in regulating many cellular functions including cell proliferation, survival, and intracellular signaling through the constant local redistribution of membrane lipids.\cite{lingwood2010lipid, smart1999caveolins,zajchowski2002lipid} 
This has major implications for protein sorting and molecular trafficking across the membrane \cite{zajchowski2002lipid}. 
Among the different molecular species involved, cholesterol is known to play a fundamental role for the correct functioning of membrane domains, regulating membrane fluidity and the structural integrity of lipid rafts. \cite{crane2004role, engberg2016affinity}.
%
%
Cholesterol depletion can lead to an increased permeability to external pathogens, signaling molecules release, and in the worst case, lipid rafts disruption with alteration of membrane thickness. This, in turn, affects the signaling pathways and can even induce programmed cell death. Cholesterol accumulation in lipid rafts is equally problematic, creating a higher sensitivity to apoptosis \cite{li2006elevated}. 
%
%
Finally, cholesterol modulates both the cell membrane local composition and its lateral molecular organisation, fluidity, and hence intercellular communication processes including endocytic pathways. \cite{hanzal2007lipid}.
One important aspect of lipid rafts - and indirectly cholesterol - is their involvement in the release and uptake of a particular class of cell-derived vesicles called extracellular vesicles (EVs). EVs are now widely accepted as nanocarriers involved in cell-cell communication \cite{huyan2020extracellular,herrmann2021extracellular,araujo2022biomedical} and have been shown to take part in many pathophysiological processes such as cancer progression and metastasis formation, cell proliferation, and stimulation of the adaptive and innate immune system \cite{becker2016extracellular,bebelman2018biogenesis}.
%
%
EVs are typically divided into two sub-classes, microvesicles (MVs) and exosomes which differ from their biogenesis pathway: budding of the membrane for MVs and endocytic pathway for exosomes \cite{kalluri2020biology}. Their respective size range is slightly different with MVs ($ 100 - 1000\; nm $) and exosomes ($30 - 150\;nm$) \cite{van2014particle,coumans2017methodological}. Since the different vesicles isolation methods rely on size-based separation, vesicles ranging from $30$ to $200\; nm$, (enriched in exosomes but also including small MVs) are often referred to as small Extracellular Vesicles (sEVs). sEVs have emerged as potential cancer biomarkers as their molecular composition reflects that of the originating cells; they are also considered optimal delivering nanocarriers as they mediate the communication between tumor and tumor-associated cells, escaping the immune response \cite{maia2018exosome}.  
%
%
This landscape is further complicated by the presence of different pathways for sEVs to deliver their molecular cargo. One of these pathways is lipid raft mediated endocytosis, where the rafts are continuously assembling/disassembling to maintain cell homeostasis and regulate vesicle trafficking \cite{mulcahy2014routes}. 
Delivery of the cargo can also occur through a mechanism initiated by some degree of fusion with the target membrane, a pathway mostly adopted by viruses. This last pathway induces a level of mixing of the sEV and target membranes which become contiguous, something recently observed with umbilical cord mesenchymal stem cells (UC-MSC) sEVs from GMP production and a model membrane containing lipid rafts \cite{perissinotto2021structural}. 
However, details on the dynamics of the interaction pathways of sEVs with target cells and on the specific role of each molecular player are still scarce and highly debated in the literature \cite{herrmann2021extracellular, russell2019biological, french2017extracellular}. 
This is related to the small size and the high heterogeneity of sEVs \cite{thery2018minimal}, as well as to the high spatial and temporal resolution required for the detection of lipid rafts dynamics. Several recent studies have investigated the role of the biophysical properties of the cell membrane on vesicle fusion and agglomeration rate, showing that the mechanical properties such as membrane curvature, fluidity and rigidity, all highly regulated by cholesterol content, can affect vesicle fusion kinetics \cite{russell2019biological, grouleff2015influence, perissinotto2021structural,caselli2021plasmon}. 
In this study, we follow up on the raft-based pathway for regulating vesicle uptake and investigate the interaction of single sEVs isolated from a triple-negative breast cancer cell line (TNBC) with model supported lipid bilayers (SLBs) with different fluidity, ordered nanodomains, and cholesterol concentration. Using Atomic Force Microscopy (AFM) in solution, we aim to explore sEVs fusion with membranes exhibiting quasi-physiological cholesterol concentration and compositions reflecting the raft structures of in vivo systems. In particular, we focus on lipid membranes with the coexistence of two lipid phases: a tightly packed and ordered state called liquid-ordered phase ($L_o$) made of sphingolipid and cholesterol molecules, coexisting with a more fluid and disordered phase called liquid-disordered ($L_d$) phase enriched with unsaturated phospholipid. The use of AFM enables us to track single EVs interacting and fusing with the membrane in situ and with nanoscale precision, and the subsequent evolution of the target membrane.
\section{Results and discussion}
\subsection{AFM of the model membrane}
\begin{figure}[H]
\includegraphics[width=0.9\textwidth]{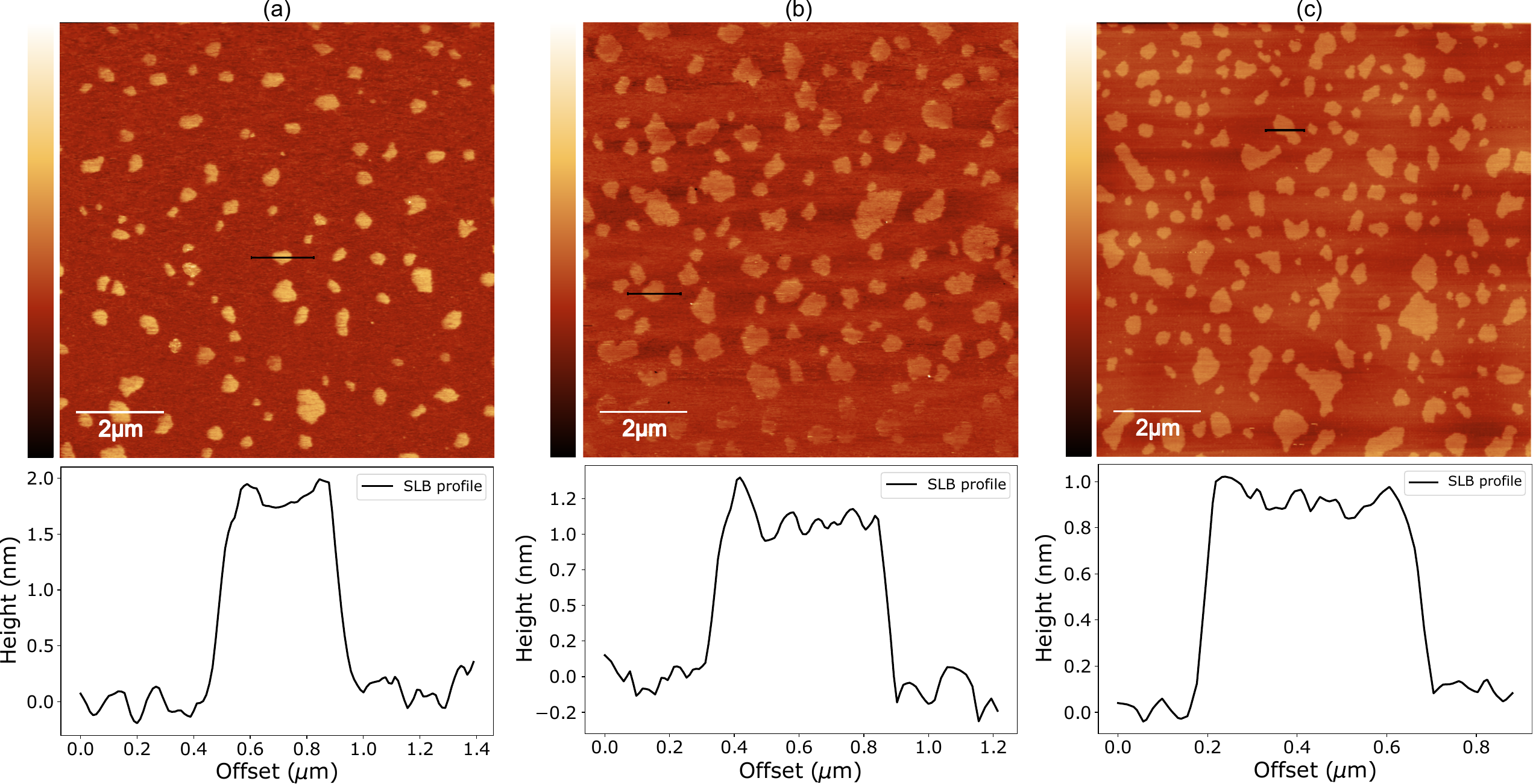}
 \caption{AFM topographic images of DOPC/SM $2:1$ (m/m) SLB with (a) $5\;mol\%$, (b) $10\;mol\%$ and (c) $17\;mol\%$ Chol. In each case a profile is shown to highlight the $L_d$ (lower) and $L_o$ (higher) domains, here acquired at room temperature in Tris $10\;mM$.}\label{fig:Figure1}
    \end{figure}

Supported lipid bilayers offer a powerful model membrane platform for studying membrane–vesicles interactions. They allow for the simultaneous analysis of the SLB morphology changes with quantitative information about the surface area of the different lipid phases, height modification, and real-time observation of the vesicle fusion process with the lipidic system. In order to gain microscopic insights into the sEV fusion process, we first performed a careful topographic analysis of the multicomponent-SLBs by means of AFM. To mimic the typical organisation of cell membranes in ’lipid raft’ microdomains, the composition adopted comprises 1,2-dioleoyl-sn-glycero-3-phosphocholine (DOPC) which is a neutral and monounsaturated phospholipid ($18:1$), sphingomyelin (SM) that represents one of the most abundant sphingolipids in the plasma membrane, and is characterized by long saturated fatty acyl chains \cite{niemela2006influence}, and cholesterol (Chol) that sterically interacts with the acyl chains of other lipids and preferentially with saturated phospholipids such as SM \cite{marquardt2016Cholesterol}. This mixture is representative of the outer membrane leaflet as it contains phosphatidylcholine and sphingomyelin as its main building blocks. To monitor the lipid phase behavior and the cholesterol dependence on SLB morphology, three cholesterol molecular concentrations have been tested: $5\; mol \%$, $10\;mol \%$ and $17\;mol \%$, with DOPC and SM kept at a fixed $2:1$ ($m/m$) ratio. In the following sections the sample composition with $17\;mol\%$ will be described in more depth and compared with a bilayer that has no sterol content. The $17\;mol \%$ Chol falls in the typical biological range of $15 - 50\;\%$ for the sterol component, and offers good reproducibility and stability when performing AFM imaging in liquid conditions \cite{sullan2010Cholesterol,redondo2012influence}.
%
%
\\
Typical examples of lipid phase separation as observed in AFM topographs are presented in Figure \ref{fig:Figure1}. 
The formation of the liquid-ordered domains is visible for all three tested conditions, although they differ in height, area, and number. The taller $L_o$ bilayer domains exhibit a maximum diameter of around $0.5\; \mu m$ with $5\;mol\%$ Chol, a value that increases with Chol and reaches around $1.5\;\mu m$ at $17\;mol\%$. The apparent number of domains changes very little with increasing cholesterol percentage, varying from an average value of $113 \pm 7.07$ to $128 \pm 5.65$ and $135 \pm 14.36$ respectively. The increase of the area occupied by $L_o$ domains is accompanied by a decrease of their relative height to with respect to the surrounding DOPC. The total area occupied by $L_o$ domains is directly related to the Chol in agreement with the theory of the preferential mixing of cholesterol with saturated lipids such as SM. This results in the increase of the area per lipid \cite{ma2016Cholesterol, mcmullen2004Cholesterol}, and a ’cholesterol-condensing effect’ on phospholipids thickening of the $L_d$ phase and a reduced height difference with the $L_o$ domains \cite{ma2016Cholesterol, hung2007condensing}. 
%
%
Lastly, it has been demonstrated that the transition temperature of a phospholipid system decreases with increasing cholesterol concentration \cite{redondo2012influence}. This explains the lower number of $L_o$ domains per scanned area observed for the $5\;mol\%$, where the transition starts at $30^\circ C$, when compared with the $10\;mol\%$ and $17\;mol\%$ where lower thermal fluctuations are required to promote the $L_o$ phase nucleation. While helpful to explain the AFM observations, a full description of the systems should also take into account the impact that a rigid substrate, which tends to stabilize the lipids and promote order (lower $T_c$) \cite{leidy2002ripples}, has on the bilayer, and the kinetics of the temperature control during sample cooling, which in turn influences the $L_o$ nucleation process and growth \cite{blanchette2008quantifying}.
%
\begin{figure}[H]
\includegraphics[width=0.9\textwidth]{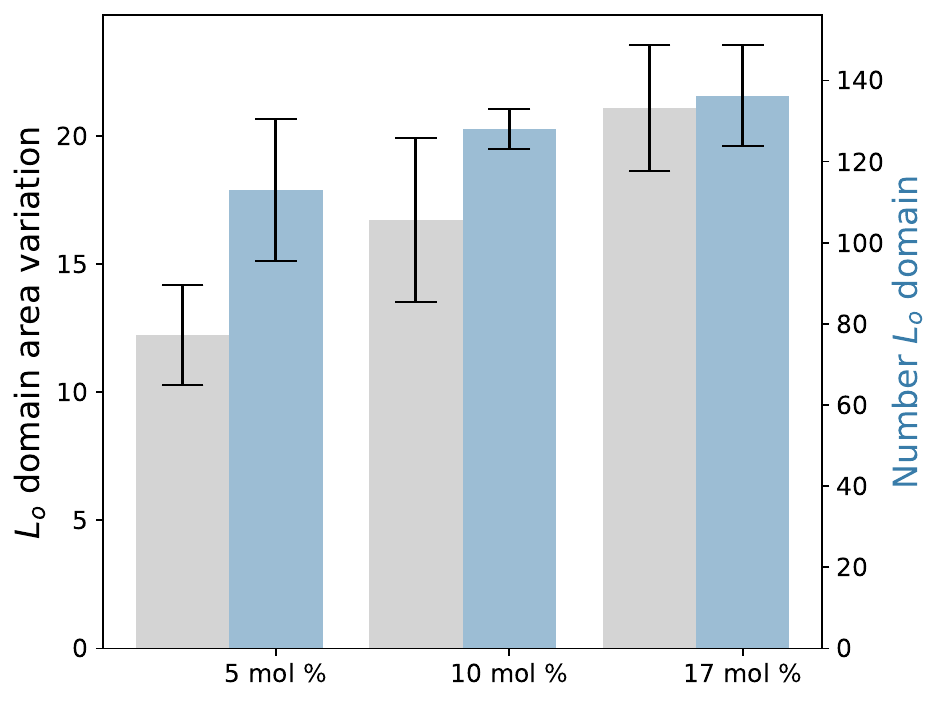}
\caption{Comparative analysis of the area and number variations of $L_o$ domains at increasing cholesterol percentage.}\label{fig:Figure2}
    \end{figure}
\subsection{Adsorption of the sEVs and local biophysical changes}
\begin{figure}[H]
\includegraphics[width=0.9\textwidth]{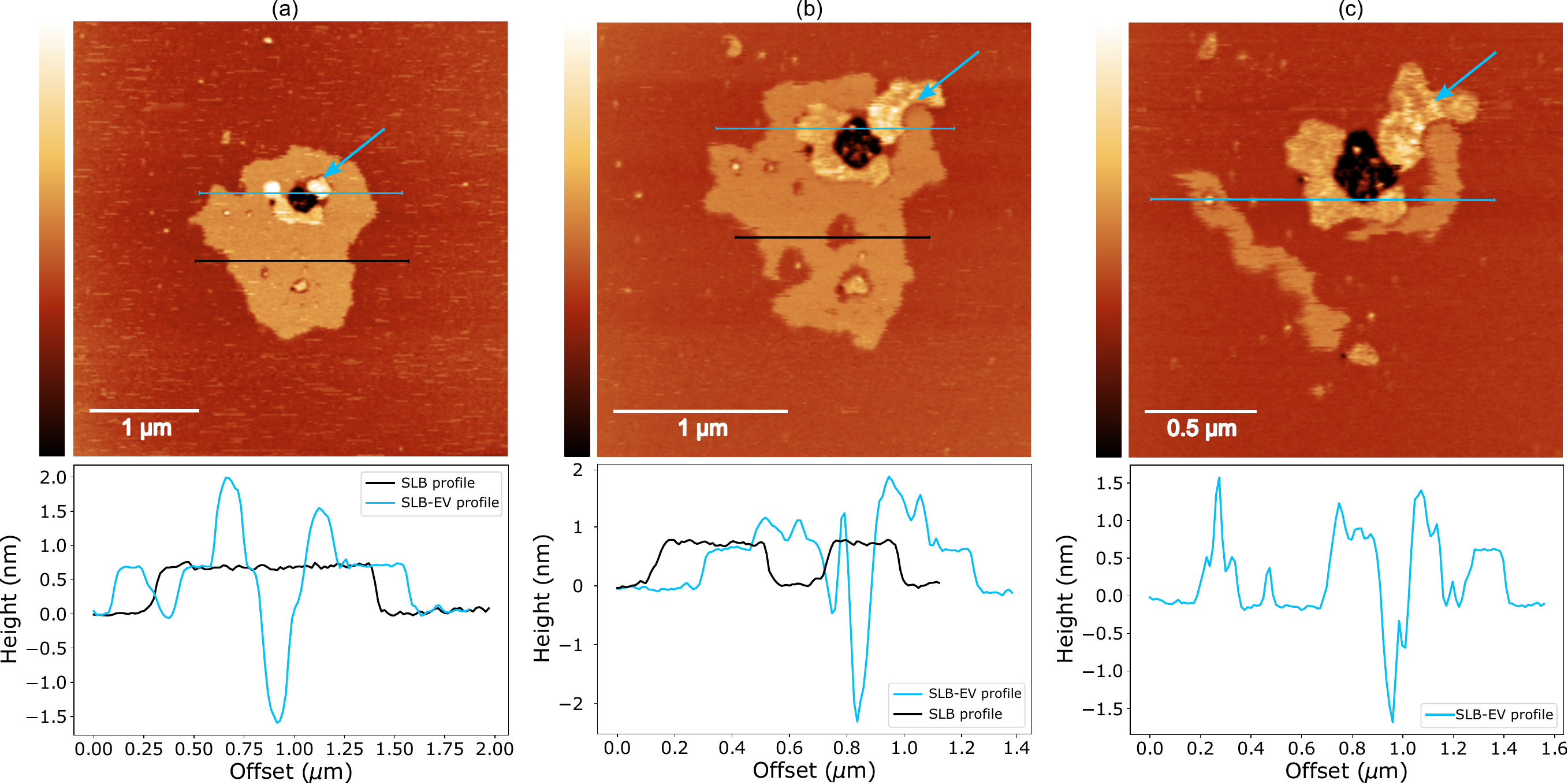}
\caption{Time-resolved AFM topographic images of EVs (MDA-MB-231 cell line) interacting with DOPC/SM $2:1$ (m/m) SLB with $17\;mol\%$ Chol with corresponding height profiles, acquired at $27\;^\circ C$ in Tris buffer $10\;mM$, with a time-lapse of 10 minutes.}\label{fig:Figure3}
    \end{figure}

With the model membrane system characterized, we then explore the interaction of sEVs isolated from the TNBC MDA-MB-231 cell line. The isolation protocol is reported in the Materials and Methods section, together with all the details of our model membrane systems. TNBC represents one of the most aggressive breast cancer subtypes, with a poor prognosis due to the absence of targetable receptors, high propensity for metastatic progression, and lack of effective chemotherapy treatments \cite{lee2019triple}. TNBC-derived sEVs have been thoroughly characterized in previous works from our group \cite{senigagliesi2022triple}. In particular, it was observed that TNBC-derived sEVs induce morphological as well as biomechanical phenotype changes in non-metastatic cancer cells toward higher aggressiveness. Here, once the model membrane has been characterized, a concentration set of sEVs are introduced to the aqueous solution and the system evolution followed in real time within few minutes. A representative AFM image of sEVs adsorption on the $17\;mol\%$ Chol membrane is shown in Figure \ref{fig:Figure3}a. 
The adsorption of sEVs induce small protrusion $\sim1\;nm$ above the height over the $L_o$ domains, accompanied by a local destabilisation of the $L_o$ region at the edges of the interaction’s site. This destabilisation appears as fluid-like regions surrounding the protrusions (blue arrows in Figure \ref{fig:Figure3}) and the formation of pores confined at the level of the outer leaflet of the supported lipid bilayer. 
%
%
Given the typical $5-6\;nm$ thickness of the membrane as measured from the SLB defects \cite{perissinotto2021structural, balgavy2001bilayer}, and the average $15.43\pm5.77\;nm$ height of sEVs when directly adsorbed on the mica substrate (according to \textcolor{blue}{Supplementary Information}) we interpret the localized protrusions as sEVs clusters whose adsorption process involves their full mixing with the SLB, and the possible molecular cargo release due to pore formation. No morphological changes were observed at the DOPC ($L_d$) level but only a local interaction with $L_o$ domains was detectable. 
%
%
To understand whether the protrusions are EV-related components or the result of the $L_o$ degradation process, a time-resolved analysis was performed to track the process evolution (Figure \ref{fig:Figure3}a-c). A drastic rearrangement of the $L_o$ domains is visible with a progressively melting into the surrounding SLB, in favor of positive growth for both the area occupied by the lipid-vesicles protrusions and SLB invaginations. After an initial step of lateral lipid redistribution with no significant morphological variations, the area occupied by $L_o$ domains progressively decreases starting from the small defects of the $L_o$ phase characterized by high curvature and evolving laterally until the melting with the $L_d$ phase expansion is completed. Simultaneously, a slight increase in the area occupied by pores and the $L_o$ phase takes place. 
%
%
The ‘melting’ effect of sEVs on planar lipid bilayer has previously been observed by our group \cite{perissinotto2021structural}, where sEVs from UC-MSC cell line were tested in the interaction with a SLB enriched with $5\;mol\%$ cholesterol. Here, sEVs lead to a dramatic fluidification of the $L_o$ phase, in contrast to the previous study where a mixing between sEVs and the $L_o$ was observed instead, with the formation of high granularity patches protruding $4\;nm$ above the SLB. These apparent differences in docking process and the resulting impact on the SLB suggests possible intrinsic differences in the EV adsorption process based on sEVs origin and cholesterol content of the target membrane. To further investigate the impact of the sEVs origins, we tested the behaviour of sEVs isolated from the UC-MSC cell line with the same target membrane containing $17\;mol\%$ Chol, resulting in qualitatively similar results to what previously reported \cite{perissinotto2021structural} (Figure 3S, \textcolor{blue}{Supplementary Information}). Given the relevance of lipid raft integrity in regulating cell proliferation, adhesion, and invasion \cite{badana2016lipid}, these results further strengthens the idea of EV potency altering the membrane properties. It also underlines the need for screening approaches that consider, other than EV’s molecular cargo and surface properties, the cell membrane molecular composition in order to be able to investigate their ability to alter the membrane properties of recipient cells, such that both faces of the interaction process can be explored.  
\subsection{sEVs interaction is regulated by lipids mobility}
\begin{figure}[H]
\includegraphics[width=0.8\textwidth]{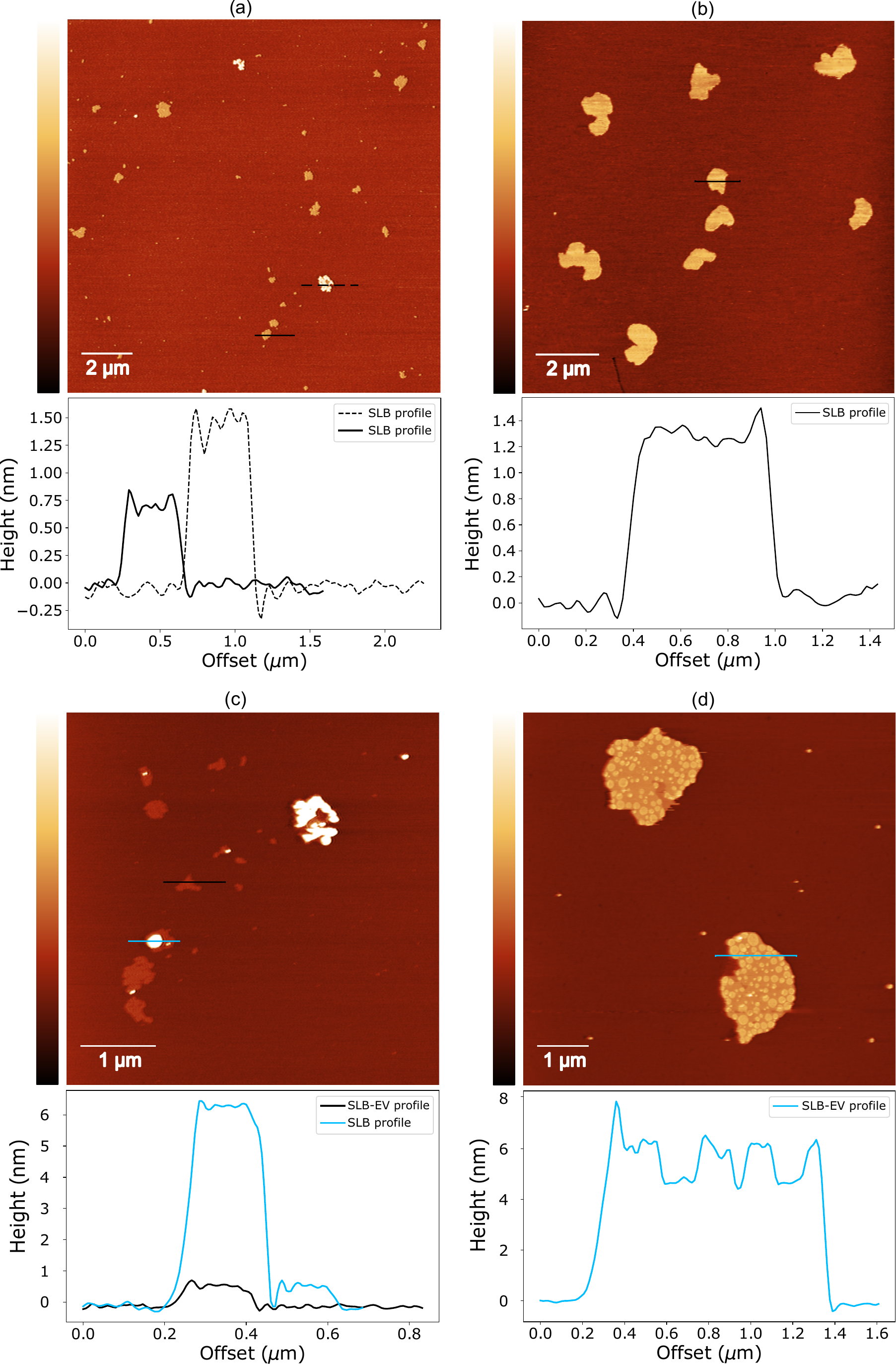}\caption{AFM topographic images of DOPC/SM and DOPC/DPPC $2:1$ (m/m) SLB before (a,b) and after (c,d) EVs (MDA-MB-231 cell line) interaction with corresponding height profiles, acquired at $27^\circ C$ in Tris buffer $10\;mM$.}\label{fig:Figure4}
    \end{figure}

The previous results highlight the importance of ordered nano-domains on the adsorption and fusion of sEVs. The well-established importance of cholesterol in modulating the emergence, stability and fate of these nano-domains makes it an obvious agent for indirectly modulating sEVs uptake in recipient cells. It is however not clear at this stage to what extent the effect is physical in terms of membrane biomechanics and fluidity or chemical through specific interactions between cholesterol and adsorbing sEVs. To further study the impact of membrane fluidity on modulating the sEVs adsorption, two control compositions with $0\;\%$ Chol content were also analysed containing either DOPC and SM $2:1$ or DOPC and DPPC $2:1$ at $27^\circ C$. In these conditions, SM domains are expected to form an ordered phase also called solid-ordered ($S_o$), characterized by a higher degree of order and less fluidity, surrounded by fluid DOPC. Similarly, DPPC domains should form highly ordered gel-phase domains within the DOPC. For both membranes, AFM imaging confirms the expectations (Figure \ref{fig:Figure4}a,b), with SM forming smaller domains covering an average percentage area of $1.17\;\%$ and protruding $1.75\;nm$ over the DOPC layer, compared to bigger DPPC domains, occupying an average $2.8\;\%$ of the membrane and with a relative height of $2\;nm$ over the surrounding DOPC. The SM $S_o$ domains are also more irregular in height that DPPC, showing two different levels at $0.75\;nm$ and $1.75\;nm$ above the DOPC layer, suggesting that the phase transition of the SM during the cooling is not uniform. Indeed, the two levels can be explained by a leaflet-by-leaflet phase transition where the SM molecules in contact with the substrate solidify first \cite{rinia2001visualizing, alessandrini2014phase}. This is also consistent with the fact that DPPC displays a highly cooperative phase transition characterized by a sharp peak at the main $T_m$ in differential scanning calorimetry whereas SM shows a single endothermic peak with a wide transition range related to the heterogeneity of the fatty acids of the lipid \cite{demetzos2008differential,nyholm2003calorimetric}.
The interactions of MDA-MB-231 sEVs with the SM $S_o$ phase is illustrated in Figure \ref{fig:Figure4}c, showing the formation of protrusions $6\;nm$ above the lipid domains. Interestingly, the sEVs clusters that co-localise with the portion of SM domains are characterized by the largest height. Moreover, the number of interaction sites per scanned area is higher compared to the membrane with $17\;mol\%$ Chol, indicating an enhanced EV interaction with the planar lipid bilayer. However, no local morphological variations can be observed over time, suggesting that the sEVs are no longer able to mix with their lipidic component with that of the SLB. A comparative experiment conducted on the DOPC/DPPC membrane displays a similar degree of order and level of saturation to the model system with SM, ruling out a chemical affinity of the sEVs with SM. Also in this case (Figure \ref{fig:Figure4}d), a specific MDA-MB-231 sEVs interaction with the ordered domains is observed. However, contrary to SM domains, a mixing with the vesicles is visible, inducing an increase of the relative height of the $S_o$ domains (profile in Figure \ref{fig:Figure4}d). This is confirmed by AFM revealing the overlapping of multiple layers and the presence of a ’vesicle-like’ morphology over the DPPC domains. To fully confirm the hypothesis of sEVs preferential mixing with high-ordered domains, two control experiments were performed using single-component SLB made of either pure DOPC or pure DPPC. The results, reported in Figure 2S of \textcolor{blue}{Supplementary Information}, confirm that sEVs do not interact with the disordered DOPC SLB, while a maximal interaction can be observed for the DPPC SLB, resulting in the SLB morphology reshaping over a larger time scale compared to the system enriched with cholesterol. These results highlight the need of lower system fluidity for the 'lipid raft' domains in order to have a fast EV adsorption process and cargo release over the SLB.  Moreover, the structural SLB modification leading to a ’lipid rafts’ fluidification further stresses the importance of the molecular orientation and packing in the recipient membrane lipids to control interaction and uptake of sEVs over time. These results pave the basis for further investigating the physicochemical mechanisms of the cell membrane, and in particular of lipid rafts as a preferential route of interaction with the sEVs.
\section{Conclusions}
The development of a multi-component SLB mimicking the 'lipid-raft' structure of cell model membranes, allowed us to study the driving forces regulating the sEVs uptake for vesicles isolated from breast cancer cell lines. Our findings, based on fast AFM topographic imaging, indicate a preferential sEV affinity for the ordered lipid raft-like domains. However, the adsorption process undergoes different pathways depending on lipid bilayer composition and fluidity. Working at the submicrometric level and performing a time-resolved analysis it was possible to identify two interaction pathways. For a fluid SLB enriched with cholesterol, the adsorption process is featured by the formation of sEV clusters protruding over the outer layer of the model system. In the same frame, a pore-opening close to the interaction site occurs, followed by a fluidification step that leads to lipid raft integrity loss. Whereas, for a rigid system without cholesterol, the adsorption pathway follows the budding-fission mechanisms \cite{liu2023kinetic}, with maximal affinity with the solid-ordered domains. This alternative mechanism is described by the fusion of the vesicles with the outer layer of the model membrane and the formation of an intermediate regular lipid phase due to full lipid mixing with the vesicles. In such a rigid system, the extent of the interaction is featured by the formation of a stable state not prone to fracture, which leads to a large-scale shape modification over time. Our study provides evidence that the degree of sEV mixing with lipids is highly regulated by the vesicle origin but also by the fluidity of the SLB. Although the lipid composition is limited to a restricted choice of lipids and cholesterol range, we believe that our results provide a strong message in light of the chemical and physical forces regulating the vesicle uptake, underling that both cell membrane composition and lateral organization must be taken into consideration to rationalize sEV interaction and cargo release in the recipient cell. Moreover, it is also evident that the side effects on lipid raft integrity are not negligible as well, as it has been demonstrated that membrane domain disruption is fundamental for the regulation of molecules trafficking across the membrane and cell survival \cite{badana2016lipid}. Furthermore, it is interesting to note that this versatile platform can be applied to study the impact of surface functionalization strategies (e.g. fusogenic proteins) on the vesicle uptake pathways \cite{verta2022generation}, but it can also be easily integrated, besides cholesterol molecules, with other lipids and proteins. In particular, the reconstitution of transmembrane proteins in the proposed model would be an innovative approach for studying transmembrane proteins localization and activity, when the planar lipid bilayer is fabricated over a pore spanning membrane \cite{teiwes2021pore, muhlenbrock2020fusion}. We foresee that, with some implementation of the model, we can develop a versatile and broadly accessible platform for the investigation the sEVs uptake pathways.
\section{Experimental Section}
\subsection{sEV isolation and characterization}
For sEV isolation, MDA-MB-231 cells ($2\cdot10^6$) were grown in a $175\; \unit{\cm}^2$ flask in DMEM (Sigma-Aldrich) with $20\;\%$ FBS (EuroClone) for 3 days. The cells were then washed two times with PBS and three times with DMEM without serum. The cells were further incubated at $37^\circ \unit{\C}$. After $24 \;\unit{\hour}$ the medium was collected and centrifuged at $300\;g$ and $4^\circ \unit{\C}$ (Allegra X-22R, Beckman Coulter) for $10\;\unit{\min}$. With a $0.22 \;\unit{\mu}\unit{\m}$ filter, the supernatant was filtered, poured into Amicon Filter Units (Ultracel-PLPLHK, $100\;\unit{\kilo\dalton}$ cutoff, Merck Millipore, UFC9100) and centrifuged at $3900 g/ 4^\circ \unit{\C}$ for $20\;\unit{\min}$ (Allegra X-22R, Beckman Coulter). The samples collected were then transferred into the polypropylene (PP) ultracentrifuge tubes (Beckman Coulter, 361623), filled with PBS and centrifuged at $ 120000\;g/4^\circ \unit{\C}$ for $2\;\unit{\hour}$ in the ultracentrifuge (70.1 Ti rotor, k-factor 36, Beckman Coulter, Brea, CA, USA). After removing the supernatant, the pellets were resuspended in $200\;\unit{\mu}\unit{\liter}$ of PBS, aliquoted, and conserved at $-20\;^\circ\unit{\C}$ until usage.
\subsection{Small unilamellar vesicles preparation}
The lipids, 1,2-dioleoyl-sn-glycero-3-phosphoCholine ($18:1$ ($\Delta9-Cis$) PC), 1,2-dipalmitoyl-sn-glycero-3-phosphoCholine (DPPC, 16:1), Sphingomyelin (brain, porcine, SM), and cholesterol (ovine wool, $>98\%$), were purchased from Avanti Polar Lipids. The single lipids, suspended in chloroform, were mixed at the desired concentration and placed under vacuum overnight. The dry film was then hydrated with TRIS buffer ($10\unit{mM}$, $pH=7.4$),  to obtain a final concentration of $1 \unit{\milli\g/\milli\l}$. The lipidic mixture was sonicated for 40 min at $45^\circ \unit{C}$ and vortexed. Lastly, the resulting solution was extruded $51$ times at $40^\circ \unit{C}$ through a membrane with $100 \unit{\nm}$ pores (PC Membranes $0.1\;\unit{\mu}\unit{\m}$, Avanti Polar Lipids).
\subsection{Supported lipid bilayers preparation}
Lipids were combined in three lipid mixtures: DOPC/SM (2:1 m/m) with Chol (5, 10, 17 \unit{mol}\%), DOPC/SM and DOPC/DPPC in a fixed molar ratio of 2:1, and lastly, DOPC and DPPC alone. The obtained extruded solution was diluted in TRIS/\ce{CaCl2} buffer to a final concentration of $0.4\;\unit{\milli\g/\milli\l}$ with $2\;\unit{mM} \;\ce{CaCl2}$. For all compositions, the vesicle fusion method was adopted as a standard procedure for planar lipid bilayer preparation. The sample was deposited on a freshly cleaved mica substrate (Nano-Tec V-1 grade, $0.15-0.21\;mm$ thickness, $10\;mm$ diameter), incubated at $50^\circ \unit{C}$ for $30\;\unit{min}$, and slowly cooled to $27^\circ \unit{C}$, then extensively washed with TRIS buffer $10\;\unit{mM}$.
\subsection{Atomic Force Microscopy imaging}
AFM was performed on commercially available microscope (Cypher ES from Asylum Research), working at $27^\circ C$ in high resolution AC mode. Sharpe nitride levers ($SNL-10$ with A geometry from Bruker Corporation) were used to perform the imaging in liquid conditions. Images were acquired at $512\times512$ pixel frames at $2.44\;\unit{Hz}$.
\section{Author contributions}
C. P., L. C., K. V. and P. P. conceived and planned the
experiments. C. P. performed the atomic force
microscopy experiments and analysed the data. V. D. contributed to atomic force microscopy measurements. V.D. and B.S. contributed to EV isolation and molecular characterization. N.T. contributed to atomic force microscopy training. C. P. and L. C. took the lead in writing the
manuscript. All authors provided critical feedback and helped
shape the research, analysis and manuscript.
\section{Conflicts of interest}
There are no conflicts to declare.
\section{Acknowledgments}
The authors wish to thank M. Gimona from Paracelsus Medical University (Salzburg, Austria) for providing the EV-UC-MSC samples. We gratefully acknowledge the Structural Biology Laboratory at Elettra-Sincrotrone Trieste S.C.p.A. for the instrumentation and constant support during the cell culture experiments. We acknowledge the Soft and Bio NanoInterfaces Laboratory at Durham University for the precious and continuous support. The authors and in particular C. P. are very grateful to CERIC-ERIC for financial funding within the framework of the INTEGRA and INTEGRA's PhD project.
\newpage
\bibliography{references}

\providecommand{\latin}[1]{#1}
\makeatletter
\providecommand{\doi}
  {\begingroup\let\do\@makeother\dospecials
  \catcode`\{=1 \catcode`\}=2 \doi@aux}
\providecommand{\doi@aux}[1]{\endgroup\texttt{#1}}
\makeatother
\providecommand*\mcitethebibliography{\thebibliography}
\csname @ifundefined\endcsname{endmcitethebibliography}
  {\let\endmcitethebibliography\endthebibliography}{}
\begin{mcitethebibliography}{48}
\providecommand*\natexlab[1]{#1}
\providecommand*\mciteSetBstSublistMode[1]{}
\providecommand*\mciteSetBstMaxWidthForm[2]{}
\providecommand*\mciteBstWouldAddEndPuncttrue
  {\def\EndOfBibitem{\unskip.}}
\providecommand*\mciteBstWouldAddEndPunctfalse
  {\let\EndOfBibitem\relax}
\providecommand*\mciteSetBstMidEndSepPunct[3]{}
\providecommand*\mciteSetBstSublistLabelBeginEnd[3]{}
\providecommand*\EndOfBibitem{}
\mciteSetBstSublistMode{f}
\mciteSetBstMaxWidthForm{subitem}{(\alph{mcitesubitemcount})}
\mciteSetBstSublistLabelBeginEnd
  {\mcitemaxwidthsubitemform\space}
  {\relax}
  {\relax}

\bibitem[Tekpli \latin{et~al.}(2013)Tekpli, Holme, Sergent, and
  Lagadic-Gossmann]{tekpli2013role}
Tekpli,~X.; Holme,~J.~A.; Sergent,~O.; Lagadic-Gossmann,~D. Role for membrane
  remodeling in cell death: implication for health and disease.
  \emph{Toxicology} \textbf{2013}, \emph{304}, 141--157\relax
\mciteBstWouldAddEndPuncttrue
\mciteSetBstMidEndSepPunct{\mcitedefaultmidpunct}
{\mcitedefaultendpunct}{\mcitedefaultseppunct}\relax
\EndOfBibitem
\bibitem[Sezgin \latin{et~al.}(2017)Sezgin, Levental, Mayor, and
  Eggeling]{sezgin2017mystery}
Sezgin,~E.; Levental,~I.; Mayor,~S.; Eggeling,~C. The mystery of membrane
  organization: composition, regulation and roles of lipid rafts. \emph{Nature
  reviews Molecular cell biology} \textbf{2017}, \emph{18}, 361--374\relax
\mciteBstWouldAddEndPuncttrue
\mciteSetBstMidEndSepPunct{\mcitedefaultmidpunct}
{\mcitedefaultendpunct}{\mcitedefaultseppunct}\relax
\EndOfBibitem
\bibitem[Simons and Sampaio(2011)Simons, and Sampaio]{simons2011membrane}
Simons,~K.; Sampaio,~J.~L. Membrane organization and lipid rafts. \emph{Cold
  Spring Harbor perspectives in biology} \textbf{2011}, \emph{3}, a004697\relax
\mciteBstWouldAddEndPuncttrue
\mciteSetBstMidEndSepPunct{\mcitedefaultmidpunct}
{\mcitedefaultendpunct}{\mcitedefaultseppunct}\relax
\EndOfBibitem
\bibitem[Lingwood and Simons(2010)Lingwood, and Simons]{lingwood2010lipid}
Lingwood,~D.; Simons,~K. Lipid rafts as a membrane-organizing principle.
  \emph{science} \textbf{2010}, \emph{327}, 46--50\relax
\mciteBstWouldAddEndPuncttrue
\mciteSetBstMidEndSepPunct{\mcitedefaultmidpunct}
{\mcitedefaultendpunct}{\mcitedefaultseppunct}\relax
\EndOfBibitem
\bibitem[Smart \latin{et~al.}(1999)Smart, Graf, McNiven, Sessa, Engelman,
  Scherer, Okamoto, and Lisanti]{smart1999caveolins}
Smart,~E.~J.; Graf,~G.~A.; McNiven,~M.~A.; Sessa,~W.~C.; Engelman,~J.~A.;
  Scherer,~P.~E.; Okamoto,~T.; Lisanti,~M.~P. Caveolins, liquid-ordered
  domains, and signal transduction. \emph{Molecular and cellular biology}
  \textbf{1999}, \emph{19}, 7289--7304\relax
\mciteBstWouldAddEndPuncttrue
\mciteSetBstMidEndSepPunct{\mcitedefaultmidpunct}
{\mcitedefaultendpunct}{\mcitedefaultseppunct}\relax
\EndOfBibitem
\bibitem[Zajchowski and Robbins(2002)Zajchowski, and
  Robbins]{zajchowski2002lipid}
Zajchowski,~L.~D.; Robbins,~S.~M. Lipid rafts and little caves:
  compartmentalized signalling in membrane microdomains. \emph{European Journal
  of Biochemistry} \textbf{2002}, \emph{269}, 737--752\relax
\mciteBstWouldAddEndPuncttrue
\mciteSetBstMidEndSepPunct{\mcitedefaultmidpunct}
{\mcitedefaultendpunct}{\mcitedefaultseppunct}\relax
\EndOfBibitem
\bibitem[Crane and Tamm(2004)Crane, and Tamm]{crane2004role}
Crane,~J.~M.; Tamm,~L.~K. Role of cholesterol in the formation and nature of
  lipid rafts in planar and spherical model membranes. \emph{Biophysical
  journal} \textbf{2004}, \emph{86}, 2965--2979\relax
\mciteBstWouldAddEndPuncttrue
\mciteSetBstMidEndSepPunct{\mcitedefaultmidpunct}
{\mcitedefaultendpunct}{\mcitedefaultseppunct}\relax
\EndOfBibitem
\bibitem[Engberg \latin{et~al.}(2016)Engberg, Hautala, Yasuda, Dehio, Murata,
  Slotte, and Nyholm]{engberg2016affinity}
Engberg,~O.; Hautala,~V.; Yasuda,~T.; Dehio,~H.; Murata,~M.; Slotte,~J.~P.;
  Nyholm,~T.~K. The affinity of cholesterol for different phospholipids affects
  lateral segregation in bilayers. \emph{Biophysical journal} \textbf{2016},
  \emph{111}, 546--556\relax
\mciteBstWouldAddEndPuncttrue
\mciteSetBstMidEndSepPunct{\mcitedefaultmidpunct}
{\mcitedefaultendpunct}{\mcitedefaultseppunct}\relax
\EndOfBibitem
\bibitem[Li \latin{et~al.}(2006)Li, Park, Ye, Kim, and Kim]{li2006elevated}
Li,~Y.~C.; Park,~M.~J.; Ye,~S.-K.; Kim,~C.-W.; Kim,~Y.-N. Elevated levels of
  cholesterol-rich lipid rafts in cancer cells are correlated with apoptosis
  sensitivity induced by cholesterol-depleting agents. \emph{The American
  journal of pathology} \textbf{2006}, \emph{168}, 1107--1118\relax
\mciteBstWouldAddEndPuncttrue
\mciteSetBstMidEndSepPunct{\mcitedefaultmidpunct}
{\mcitedefaultendpunct}{\mcitedefaultseppunct}\relax
\EndOfBibitem
\bibitem[Hanzal-Bayer and Hancock(2007)Hanzal-Bayer, and
  Hancock]{hanzal2007lipid}
Hanzal-Bayer,~M.~F.; Hancock,~J.~F. Lipid rafts and membrane traffic.
  \emph{FEBS letters} \textbf{2007}, \emph{581}, 2098--2104\relax
\mciteBstWouldAddEndPuncttrue
\mciteSetBstMidEndSepPunct{\mcitedefaultmidpunct}
{\mcitedefaultendpunct}{\mcitedefaultseppunct}\relax
\EndOfBibitem
\bibitem[Huyan \latin{et~al.}(2020)Huyan, Li, Peng, Chen, Yang, Zhang, and
  Li]{huyan2020extracellular}
Huyan,~T.; Li,~H.; Peng,~H.; Chen,~J.; Yang,~R.; Zhang,~W.; Li,~Q.
  Extracellular vesicles--advanced nanocarriers in cancer therapy: progress and
  achievements. \emph{International journal of nanomedicine} \textbf{2020},
  6485--6502\relax
\mciteBstWouldAddEndPuncttrue
\mciteSetBstMidEndSepPunct{\mcitedefaultmidpunct}
{\mcitedefaultendpunct}{\mcitedefaultseppunct}\relax
\EndOfBibitem
\bibitem[Herrmann \latin{et~al.}(2021)Herrmann, Wood, and
  Fuhrmann]{herrmann2021extracellular}
Herrmann,~I.~K.; Wood,~M. J.~A.; Fuhrmann,~G. Extracellular vesicles as a
  next-generation drug delivery platform. \emph{Nature nanotechnology}
  \textbf{2021}, \emph{16}, 748--759\relax
\mciteBstWouldAddEndPuncttrue
\mciteSetBstMidEndSepPunct{\mcitedefaultmidpunct}
{\mcitedefaultendpunct}{\mcitedefaultseppunct}\relax
\EndOfBibitem
\bibitem[Araujo-Abad \latin{et~al.}(2022)Araujo-Abad, Saceda, and
  de~Juan~Romero]{araujo2022biomedical}
Araujo-Abad,~S.; Saceda,~M.; de~Juan~Romero,~C. Biomedical application of small
  extracellular vesicles in cancer treatment. \emph{Advanced drug delivery
  reviews} \textbf{2022}, 114117\relax
\mciteBstWouldAddEndPuncttrue
\mciteSetBstMidEndSepPunct{\mcitedefaultmidpunct}
{\mcitedefaultendpunct}{\mcitedefaultseppunct}\relax
\EndOfBibitem
\bibitem[Becker \latin{et~al.}(2016)Becker, Thakur, Weiss, Kim, Peinado, and
  Lyden]{becker2016extracellular}
Becker,~A.; Thakur,~B.~K.; Weiss,~J.~M.; Kim,~H.~S.; Peinado,~H.; Lyden,~D.
  Extracellular vesicles in cancer: cell-to-cell mediators of metastasis.
  \emph{Cancer cell} \textbf{2016}, \emph{30}, 836--848\relax
\mciteBstWouldAddEndPuncttrue
\mciteSetBstMidEndSepPunct{\mcitedefaultmidpunct}
{\mcitedefaultendpunct}{\mcitedefaultseppunct}\relax
\EndOfBibitem
\bibitem[Bebelman \latin{et~al.}(2018)Bebelman, Smit, Pegtel, and
  Baglio]{bebelman2018biogenesis}
Bebelman,~M.~P.; Smit,~M.~J.; Pegtel,~D.~M.; Baglio,~S.~R. Biogenesis and
  function of extracellular vesicles in cancer. \emph{Pharmacology \&
  therapeutics} \textbf{2018}, \emph{188}, 1--11\relax
\mciteBstWouldAddEndPuncttrue
\mciteSetBstMidEndSepPunct{\mcitedefaultmidpunct}
{\mcitedefaultendpunct}{\mcitedefaultseppunct}\relax
\EndOfBibitem
\bibitem[Kalluri and LeBleu(2020)Kalluri, and LeBleu]{kalluri2020biology}
Kalluri,~R.; LeBleu,~V.~S. The biology, function, and biomedical applications
  of exosomes. \emph{Science} \textbf{2020}, \emph{367}, eaau6977\relax
\mciteBstWouldAddEndPuncttrue
\mciteSetBstMidEndSepPunct{\mcitedefaultmidpunct}
{\mcitedefaultendpunct}{\mcitedefaultseppunct}\relax
\EndOfBibitem
\bibitem[Van~der Pol \latin{et~al.}(2014)Van~der Pol, Coumans, Grootemaat,
  Gardiner, Sargent, Harrison, Sturk, Van~Leeuwen, and
  Nieuwland]{van2014particle}
Van~der Pol,~E.; Coumans,~F.; Grootemaat,~A.; Gardiner,~C.; Sargent,~I.~L.;
  Harrison,~P.; Sturk,~A.; Van~Leeuwen,~T.; Nieuwland,~R. Particle size
  distribution of exosomes and microvesicles determined by transmission
  electron microscopy, flow cytometry, nanoparticle tracking analysis, and
  resistive pulse sensing. \emph{Journal of Thrombosis and Haemostasis}
  \textbf{2014}, \emph{12}, 1182--1192\relax
\mciteBstWouldAddEndPuncttrue
\mciteSetBstMidEndSepPunct{\mcitedefaultmidpunct}
{\mcitedefaultendpunct}{\mcitedefaultseppunct}\relax
\EndOfBibitem
\bibitem[Coumans \latin{et~al.}(2017)Coumans, Brisson, Buzas, Dignat-George,
  Drees, El-Andaloussi, Emanueli, Gasecka, Hendrix, Hill, \latin{et~al.}
  others]{coumans2017methodological}
Coumans,~F.~A.; Brisson,~A.~R.; Buzas,~E.~I.; Dignat-George,~F.; Drees,~E.~E.;
  El-Andaloussi,~S.; Emanueli,~C.; Gasecka,~A.; Hendrix,~A.; Hill,~A.~F.,
  \latin{et~al.}  Methodological guidelines to study extracellular vesicles.
  \emph{Circulation research} \textbf{2017}, \emph{120}, 1632--1648\relax
\mciteBstWouldAddEndPuncttrue
\mciteSetBstMidEndSepPunct{\mcitedefaultmidpunct}
{\mcitedefaultendpunct}{\mcitedefaultseppunct}\relax
\EndOfBibitem
\bibitem[Maia \latin{et~al.}(2018)Maia, Caja, Strano~Moraes, Couto, and
  Costa-Silva]{maia2018exosome}
Maia,~J.; Caja,~S.; Strano~Moraes,~M.~C.; Couto,~N.; Costa-Silva,~B.
  Exosome-based cell-cell communication in the tumor microenvironment.
  \emph{Frontiers in cell and developmental biology} \textbf{2018}, \emph{6},
  18\relax
\mciteBstWouldAddEndPuncttrue
\mciteSetBstMidEndSepPunct{\mcitedefaultmidpunct}
{\mcitedefaultendpunct}{\mcitedefaultseppunct}\relax
\EndOfBibitem
\bibitem[Mulcahy \latin{et~al.}(2014)Mulcahy, Pink, and
  Carter]{mulcahy2014routes}
Mulcahy,~L.~A.; Pink,~R.~C.; Carter,~D. R.~F. Routes and mechanisms of
  extracellular vesicle uptake. \emph{Journal of extracellular vesicles}
  \textbf{2014}, \emph{3}, 24641\relax
\mciteBstWouldAddEndPuncttrue
\mciteSetBstMidEndSepPunct{\mcitedefaultmidpunct}
{\mcitedefaultendpunct}{\mcitedefaultseppunct}\relax
\EndOfBibitem
\bibitem[Perissinotto \latin{et~al.}(2021)Perissinotto, Rondelli, Senigagliesi,
  Brocca, Alm{\'a}sy, Botty{\'a}n, Merkel, Amenitsch, Sartori, Pachler,
  \latin{et~al.} others]{perissinotto2021structural}
Perissinotto,~F.; Rondelli,~V.; Senigagliesi,~B.; Brocca,~P.; Alm{\'a}sy,~L.;
  Botty{\'a}n,~L.; Merkel,~D.~G.; Amenitsch,~H.; Sartori,~B.; Pachler,~K.,
  \latin{et~al.}  Structural insights into fusion mechanisms of small
  extracellular vesicles with model plasma membranes. \emph{Nanoscale}
  \textbf{2021}, \emph{13}, 5224--5233\relax
\mciteBstWouldAddEndPuncttrue
\mciteSetBstMidEndSepPunct{\mcitedefaultmidpunct}
{\mcitedefaultendpunct}{\mcitedefaultseppunct}\relax
\EndOfBibitem
\bibitem[Russell \latin{et~al.}(2019)Russell, Sneider, Witwer, Bergese,
  Bhattacharyya, Cocks, Cocucci, Erdbr{\"u}gger, Falcon-Perez, Freeman,
  \latin{et~al.} others]{russell2019biological}
Russell,~A.~E.; Sneider,~A.; Witwer,~K.~W.; Bergese,~P.; Bhattacharyya,~S.~N.;
  Cocks,~A.; Cocucci,~E.; Erdbr{\"u}gger,~U.; Falcon-Perez,~J.~M.;
  Freeman,~D.~W., \latin{et~al.}  Biological membranes in EV biogenesis,
  stability, uptake, and cargo transfer: an ISEV position paper arising from
  the ISEV membranes and EVs workshop. \emph{Journal of Extracellular Vesicles}
  \textbf{2019}, \emph{8}, 1684862\relax
\mciteBstWouldAddEndPuncttrue
\mciteSetBstMidEndSepPunct{\mcitedefaultmidpunct}
{\mcitedefaultendpunct}{\mcitedefaultseppunct}\relax
\EndOfBibitem
\bibitem[French \latin{et~al.}(2017)French, Antonyak, and
  Cerione]{french2017extracellular}
French,~K.~C.; Antonyak,~M.~A.; Cerione,~R.~A. Extracellular vesicle docking at
  the cellular port: Extracellular vesicle binding and uptake. Seminars in cell
  \& developmental biology. 2017; pp 48--55\relax
\mciteBstWouldAddEndPuncttrue
\mciteSetBstMidEndSepPunct{\mcitedefaultmidpunct}
{\mcitedefaultendpunct}{\mcitedefaultseppunct}\relax
\EndOfBibitem
\bibitem[Th{\'e}ry \latin{et~al.}(2018)Th{\'e}ry, Witwer, Aikawa, Alcaraz,
  Anderson, Andriantsitohaina, Antoniou, Arab, Archer, Atkin-Smith,
  \latin{et~al.} others]{thery2018minimal}
Th{\'e}ry,~C.; Witwer,~K.~W.; Aikawa,~E.; Alcaraz,~M.~J.; Anderson,~J.~D.;
  Andriantsitohaina,~R.; Antoniou,~A.; Arab,~T.; Archer,~F.;
  Atkin-Smith,~G.~K., \latin{et~al.}  Minimal information for studies of
  extracellular vesicles 2018 (MISEV2018): a position statement of the
  International Society for Extracellular Vesicles and update of the MISEV2014
  guidelines. \emph{Journal of extracellular vesicles} \textbf{2018}, \emph{7},
  1535750\relax
\mciteBstWouldAddEndPuncttrue
\mciteSetBstMidEndSepPunct{\mcitedefaultmidpunct}
{\mcitedefaultendpunct}{\mcitedefaultseppunct}\relax
\EndOfBibitem
\bibitem[Grouleff \latin{et~al.}(2015)Grouleff, Irudayam, Skeby, and
  Schi{\o}tt]{grouleff2015influence}
Grouleff,~J.; Irudayam,~S.~J.; Skeby,~K.~K.; Schi{\o}tt,~B. The influence of
  cholesterol on membrane protein structure, function, and dynamics studied by
  molecular dynamics simulations. \emph{Biochimica et Biophysica Acta
  (BBA)-Biomembranes} \textbf{2015}, \emph{1848}, 1783--1795\relax
\mciteBstWouldAddEndPuncttrue
\mciteSetBstMidEndSepPunct{\mcitedefaultmidpunct}
{\mcitedefaultendpunct}{\mcitedefaultseppunct}\relax
\EndOfBibitem
\bibitem[Caselli \latin{et~al.}(2021)Caselli, Ridolfi, Cardellini, Sharpnack,
  Paolini, Brucale, Valle, Montis, Bergese, and Berti]{caselli2021plasmon}
Caselli,~L.; Ridolfi,~A.; Cardellini,~J.; Sharpnack,~L.; Paolini,~L.;
  Brucale,~M.; Valle,~F.; Montis,~C.; Bergese,~P.; Berti,~D. A plasmon-based
  nanoruler to probe the mechanical properties of synthetic and biogenic
  nanosized lipid vesicles. \emph{Nanoscale Horizons} \textbf{2021}, \emph{6},
  543--550\relax
\mciteBstWouldAddEndPuncttrue
\mciteSetBstMidEndSepPunct{\mcitedefaultmidpunct}
{\mcitedefaultendpunct}{\mcitedefaultseppunct}\relax
\EndOfBibitem
\bibitem[Niemel{\"a} \latin{et~al.}(2006)Niemel{\"a}, Hyv{\"o}nen, and
  Vattulainen]{niemela2006influence}
Niemel{\"a},~P.~S.; Hyv{\"o}nen,~M.~T.; Vattulainen,~I. Influence of chain
  length and unsaturation on sphingomyelin bilayers. \emph{Biophysical journal}
  \textbf{2006}, \emph{90}, 851--863\relax
\mciteBstWouldAddEndPuncttrue
\mciteSetBstMidEndSepPunct{\mcitedefaultmidpunct}
{\mcitedefaultendpunct}{\mcitedefaultseppunct}\relax
\EndOfBibitem
\bibitem[Marquardt \latin{et~al.}(2016)Marquardt, Ku{\v{c}}erka, Wassall,
  Harroun, and Katsaras]{marquardt2016Cholesterol}
Marquardt,~D.; Ku{\v{c}}erka,~N.; Wassall,~S.~R.; Harroun,~T.~A.; Katsaras,~J.
  Cholesterol's location in lipid bilayers. \emph{Chemistry and Physics of
  Lipids} \textbf{2016}, \emph{199}, 17--25\relax
\mciteBstWouldAddEndPuncttrue
\mciteSetBstMidEndSepPunct{\mcitedefaultmidpunct}
{\mcitedefaultendpunct}{\mcitedefaultseppunct}\relax
\EndOfBibitem
\bibitem[Sullan \latin{et~al.}(2010)Sullan, Li, Hao, Walker, and
  Zou]{sullan2010Cholesterol}
Sullan,~R. M.~A.; Li,~J.~K.; Hao,~C.; Walker,~G.~C.; Zou,~S.
  Cholesterol-dependent nanomechanical stability of phase-segregated
  multicomponent lipid bilayers. \emph{Biophysical journal} \textbf{2010},
  \emph{99}, 507--516\relax
\mciteBstWouldAddEndPuncttrue
\mciteSetBstMidEndSepPunct{\mcitedefaultmidpunct}
{\mcitedefaultendpunct}{\mcitedefaultseppunct}\relax
\EndOfBibitem
\bibitem[Redondo-Morata \latin{et~al.}(2012)Redondo-Morata, Giannotti, and
  Sanz]{redondo2012influence}
Redondo-Morata,~L.; Giannotti,~M.~I.; Sanz,~F. Influence of cholesterol on the
  phase transition of lipid bilayers: a temperature-controlled force
  spectroscopy study. \emph{Langmuir} \textbf{2012}, \emph{28},
  12851--12860\relax
\mciteBstWouldAddEndPuncttrue
\mciteSetBstMidEndSepPunct{\mcitedefaultmidpunct}
{\mcitedefaultendpunct}{\mcitedefaultseppunct}\relax
\EndOfBibitem
\bibitem[Ma \latin{et~al.}(2016)Ma, Ghosh, DiLena, Bera, Lurio, Parikh, and
  Sinha]{ma2016Cholesterol}
Ma,~Y.; Ghosh,~S.~K.; DiLena,~D.~A.; Bera,~S.; Lurio,~L.~B.; Parikh,~A.~N.;
  Sinha,~S.~K. Cholesterol partition and condensing effect in phase-separated
  ternary mixture lipid multilayers. \emph{Biophysical journal} \textbf{2016},
  \emph{110}, 1355--1366\relax
\mciteBstWouldAddEndPuncttrue
\mciteSetBstMidEndSepPunct{\mcitedefaultmidpunct}
{\mcitedefaultendpunct}{\mcitedefaultseppunct}\relax
\EndOfBibitem
\bibitem[McMullen \latin{et~al.}(2004)McMullen, Lewis, and
  McElhaney]{mcmullen2004Cholesterol}
McMullen,~T.~P.; Lewis,~R.~N.; McElhaney,~R.~N. Cholesterol--phospholipid
  interactions, the liquid-ordered phase and lipid rafts in model and
  biological membranes. \emph{Current opinion in colloid \& interface science}
  \textbf{2004}, \emph{8}, 459--468\relax
\mciteBstWouldAddEndPuncttrue
\mciteSetBstMidEndSepPunct{\mcitedefaultmidpunct}
{\mcitedefaultendpunct}{\mcitedefaultseppunct}\relax
\EndOfBibitem
\bibitem[Hung \latin{et~al.}(2007)Hung, Lee, Chen, and
  Huang]{hung2007condensing}
Hung,~W.-C.; Lee,~M.-T.; Chen,~F.-Y.; Huang,~H.~W. The condensing effect of
  cholesterol in lipid bilayers. \emph{Biophysical journal} \textbf{2007},
  \emph{92}, 3960--3967\relax
\mciteBstWouldAddEndPuncttrue
\mciteSetBstMidEndSepPunct{\mcitedefaultmidpunct}
{\mcitedefaultendpunct}{\mcitedefaultseppunct}\relax
\EndOfBibitem
\bibitem[Leidy \latin{et~al.}(2002)Leidy, Kaasgaard, Crowe, Mouritsen, and
  J{\o}rgensen]{leidy2002ripples}
Leidy,~C.; Kaasgaard,~T.; Crowe,~J.~H.; Mouritsen,~O.~G.; J{\o}rgensen,~K.
  Ripples and the formation of anisotropic lipid domains: imaging two-component
  supported double bilayers by atomic force microscopy. \emph{Biophysical
  journal} \textbf{2002}, \emph{83}, 2625--2633\relax
\mciteBstWouldAddEndPuncttrue
\mciteSetBstMidEndSepPunct{\mcitedefaultmidpunct}
{\mcitedefaultendpunct}{\mcitedefaultseppunct}\relax
\EndOfBibitem
\bibitem[Blanchette \latin{et~al.}(2008)Blanchette, Orme, Ratto, and
  Longo]{blanchette2008quantifying}
Blanchette,~C.~D.; Orme,~C.~A.; Ratto,~T.~V.; Longo,~M.~L. Quantifying growth
  of symmetric and asymmetric lipid bilayer domains. \emph{Langmuir}
  \textbf{2008}, \emph{24}, 1219--1224\relax
\mciteBstWouldAddEndPuncttrue
\mciteSetBstMidEndSepPunct{\mcitedefaultmidpunct}
{\mcitedefaultendpunct}{\mcitedefaultseppunct}\relax
\EndOfBibitem
\bibitem[Lee \latin{et~al.}(2019)Lee, Kuo, Ho, and Huang]{lee2019triple}
Lee,~K.-L.; Kuo,~Y.-C.; Ho,~Y.-S.; Huang,~Y.-H. Triple-negative breast cancer:
  current understanding and future therapeutic breakthrough targeting cancer
  stemness. \emph{Cancers} \textbf{2019}, \emph{11}, 1334\relax
\mciteBstWouldAddEndPuncttrue
\mciteSetBstMidEndSepPunct{\mcitedefaultmidpunct}
{\mcitedefaultendpunct}{\mcitedefaultseppunct}\relax
\EndOfBibitem
\bibitem[Senigagliesi \latin{et~al.}(2022)Senigagliesi, Samperi, Cefarin, Gneo,
  Petrosino, Apollonio, Caponnetto, Sgarra, Collavin, Cesselli, \latin{et~al.}
  others]{senigagliesi2022triple}
Senigagliesi,~B.; Samperi,~G.; Cefarin,~N.; Gneo,~L.; Petrosino,~S.;
  Apollonio,~M.; Caponnetto,~F.; Sgarra,~R.; Collavin,~L.; Cesselli,~D.,
  \latin{et~al.}  Triple negative breast cancer-derived small extracellular
  vesicles as modulator of biomechanics in target cells. \emph{Nanomedicine:
  Nanotechnology, Biology and Medicine} \textbf{2022}, \emph{44}, 102582\relax
\mciteBstWouldAddEndPuncttrue
\mciteSetBstMidEndSepPunct{\mcitedefaultmidpunct}
{\mcitedefaultendpunct}{\mcitedefaultseppunct}\relax
\EndOfBibitem
\bibitem[Balgav{\`y} \latin{et~al.}(2001)Balgav{\`y}, Dubni{\v{c}}kov{\'a},
  Ku{\v{c}}erka, Kiselev, Yaradaikin, and Uhr{\i}kov{\'a}]{balgavy2001bilayer}
Balgav{\`y},~P.; Dubni{\v{c}}kov{\'a},~M.; Ku{\v{c}}erka,~N.; Kiselev,~M.~A.;
  Yaradaikin,~S.~P.; Uhr{\i}kov{\'a},~D. Bilayer thickness and lipid interface
  area in unilamellar extruded 1, 2-diacylphosphatidylcholine liposomes: a
  small-angle neutron scattering study. \emph{Biochimica et Biophysica Acta
  (BBA)-Biomembranes} \textbf{2001}, \emph{1512}, 40--52\relax
\mciteBstWouldAddEndPuncttrue
\mciteSetBstMidEndSepPunct{\mcitedefaultmidpunct}
{\mcitedefaultendpunct}{\mcitedefaultseppunct}\relax
\EndOfBibitem
\bibitem[Badana \latin{et~al.}(2016)Badana, Chintala, Varikuti, Pudi, Kumari,
  Kappala, and Malla]{badana2016lipid}
Badana,~A.; Chintala,~M.; Varikuti,~G.; Pudi,~N.; Kumari,~S.; Kappala,~V.~R.;
  Malla,~R.~R. Lipid raft integrity is required for survival of triple negative
  breast cancer cells. \emph{Journal of breast cancer} \textbf{2016},
  \emph{19}, 372--384\relax
\mciteBstWouldAddEndPuncttrue
\mciteSetBstMidEndSepPunct{\mcitedefaultmidpunct}
{\mcitedefaultendpunct}{\mcitedefaultseppunct}\relax
\EndOfBibitem
\bibitem[Rinia \latin{et~al.}(2001)Rinia, Snel, van~der Eerden, and
  de~Kruijff]{rinia2001visualizing}
Rinia,~H.~A.; Snel,~M.~M.; van~der Eerden,~J.~P.; de~Kruijff,~B. Visualizing
  detergent resistant domains in model membranes with atomic force microscopy.
  \emph{Febs Letters} \textbf{2001}, \emph{501}, 92--96\relax
\mciteBstWouldAddEndPuncttrue
\mciteSetBstMidEndSepPunct{\mcitedefaultmidpunct}
{\mcitedefaultendpunct}{\mcitedefaultseppunct}\relax
\EndOfBibitem
\bibitem[Alessandrini and Facci(2014)Alessandrini, and
  Facci]{alessandrini2014phase}
Alessandrini,~A.; Facci,~P. Phase transitions in supported lipid bilayers
  studied by AFM. \emph{Soft matter} \textbf{2014}, \emph{10}, 7145--7164\relax
\mciteBstWouldAddEndPuncttrue
\mciteSetBstMidEndSepPunct{\mcitedefaultmidpunct}
{\mcitedefaultendpunct}{\mcitedefaultseppunct}\relax
\EndOfBibitem
\bibitem[Demetzos(2008)]{demetzos2008differential}
Demetzos,~C. Differential scanning calorimetry (DSC): a tool to study the
  thermal behavior of lipid bilayers and liposomal stability. \emph{Journal of
  liposome research} \textbf{2008}, \emph{18}, 159--173\relax
\mciteBstWouldAddEndPuncttrue
\mciteSetBstMidEndSepPunct{\mcitedefaultmidpunct}
{\mcitedefaultendpunct}{\mcitedefaultseppunct}\relax
\EndOfBibitem
\bibitem[Nyholm \latin{et~al.}(2003)Nyholm, Nylund, and
  Slotte]{nyholm2003calorimetric}
Nyholm,~T.~K.; Nylund,~M.; Slotte,~J.~P. A calorimetric study of binary
  mixtures of dihydrosphingomyelin and sterols, sphingomyelin, or
  phosphatidylcholine. \emph{Biophysical journal} \textbf{2003}, \emph{84},
  3138--3146\relax
\mciteBstWouldAddEndPuncttrue
\mciteSetBstMidEndSepPunct{\mcitedefaultmidpunct}
{\mcitedefaultendpunct}{\mcitedefaultseppunct}\relax
\EndOfBibitem
\bibitem[Liu \latin{et~al.}(2023)Liu, Duan, and Wang]{liu2023kinetic}
Liu,~L.; Duan,~C.; Wang,~R. Kinetic Pathway and Micromechanics of Vesicle
  Fusion/Fission. 2023\relax
\mciteBstWouldAddEndPuncttrue
\mciteSetBstMidEndSepPunct{\mcitedefaultmidpunct}
{\mcitedefaultendpunct}{\mcitedefaultseppunct}\relax
\EndOfBibitem
\bibitem[Verta \latin{et~al.}(2022)Verta, Grange, Skovronova, Tanzi, Peruzzi,
  Deregibus, Camussi, and Bussolati]{verta2022generation}
Verta,~R.; Grange,~C.; Skovronova,~R.; Tanzi,~A.; Peruzzi,~L.;
  Deregibus,~M.~C.; Camussi,~G.; Bussolati,~B. Generation of
  Spike-Extracellular Vesicles (S-EVs) as a Tool to Mimic SARS-CoV-2
  Interaction with Host Cells. \emph{Cells} \textbf{2022}, \emph{11}, 146\relax
\mciteBstWouldAddEndPuncttrue
\mciteSetBstMidEndSepPunct{\mcitedefaultmidpunct}
{\mcitedefaultendpunct}{\mcitedefaultseppunct}\relax
\EndOfBibitem
\bibitem[Teiwes \latin{et~al.}(2021)Teiwes, Mey, Baumann, Strieker, Unkelbach,
  and Steinem]{teiwes2021pore}
Teiwes,~N.~K.; Mey,~I.; Baumann,~P.~C.; Strieker,~L.; Unkelbach,~U.;
  Steinem,~C. Pore-Spanning Plasma Membranes Derived from Giant Plasma Membrane
  Vesicles. \emph{ACS Applied Materials \& Interfaces} \textbf{2021},
  \emph{13}, 25805--25812\relax
\mciteBstWouldAddEndPuncttrue
\mciteSetBstMidEndSepPunct{\mcitedefaultmidpunct}
{\mcitedefaultendpunct}{\mcitedefaultseppunct}\relax
\EndOfBibitem
\bibitem[M{\"u}hlenbrock \latin{et~al.}(2020)M{\"u}hlenbrock, Herwig, Vuong,
  Mey, and Steinem]{muhlenbrock2020fusion}
M{\"u}hlenbrock,~P.; Herwig,~K.; Vuong,~L.; Mey,~I.; Steinem,~C. Fusion pore
  formation observed during SNARE-mediated vesicle fusion with pore-spanning
  membranes. \emph{Biophysical Journal} \textbf{2020}, \emph{119},
  151--161\relax
\mciteBstWouldAddEndPuncttrue
\mciteSetBstMidEndSepPunct{\mcitedefaultmidpunct}
{\mcitedefaultendpunct}{\mcitedefaultseppunct}\relax
\EndOfBibitem
\end{mcitethebibliography}


\providecommand{\latin}[1]{#1}
\makeatletter
\providecommand{\doi}
  {\begingroup\let\do\@makeother\dospecials
  \catcode`\{=1 \catcode`\}=2 \doi@aux}
\providecommand{\doi@aux}[1]{\endgroup\texttt{#1}}
\makeatother
\providecommand*\mcitethebibliography{\thebibliography}
\csname @ifundefined\endcsname{endmcitethebibliography}
  {\let\endmcitethebibliography\endthebibliography}{}
\begin{mcitethebibliography}{4}
\providecommand*\natexlab[1]{#1}
\providecommand*\mciteSetBstSublistMode[1]{}
\providecommand*\mciteSetBstMaxWidthForm[2]{}
\providecommand*\mciteBstWouldAddEndPuncttrue
  {\def\EndOfBibitem{\unskip.}}
\providecommand*\mciteBstWouldAddEndPunctfalse
  {\let\EndOfBibitem\relax}
\providecommand*\mciteSetBstMidEndSepPunct[3]{}
\providecommand*\mciteSetBstSublistLabelBeginEnd[3]{}
\providecommand*\EndOfBibitem{}
\mciteSetBstSublistMode{f}
\mciteSetBstMaxWidthForm{subitem}{(\alph{mcitesubitemcount})}
\mciteSetBstSublistLabelBeginEnd
  {\mcitemaxwidthsubitemform\space}
  {\relax}
  {\relax}

\bibitem[Perissinotto \latin{et~al.}(2020)Perissinotto, Senigagliesi, Vaccari,
  Pachetti, D’Amico, Amenitsch, Sartori, Pachler, Mayr, Gimona,
  \latin{et~al.} others]{perissinotto2020multi}
Perissinotto,~F.; Senigagliesi,~B.; Vaccari,~L.; Pachetti,~M.; D’Amico,~F.;
  Amenitsch,~H.; Sartori,~B.; Pachler,~K.; Mayr,~M.; Gimona,~M., \latin{et~al.}
   \emph{Advances in Biomembranes and Lipid Self-Assembly}; Elsevier, 2020;
  Vol.~32; pp 157--177\relax
\mciteBstWouldAddEndPuncttrue
\mciteSetBstMidEndSepPunct{\mcitedefaultmidpunct}
{\mcitedefaultendpunct}{\mcitedefaultseppunct}\relax
\EndOfBibitem
\bibitem[Ridolfi \latin{et~al.}(2020)Ridolfi, Brucale, Montis, Caselli,
  Paolini, Borup, Boysen, Loria, van Herwijnen, Kleinjan, \latin{et~al.}
  others]{ridolfi2020afm}
Ridolfi,~A.; Brucale,~M.; Montis,~C.; Caselli,~L.; Paolini,~L.; Borup,~A.;
  Boysen,~A.~T.; Loria,~F.; van Herwijnen,~M.~J.; Kleinjan,~M., \latin{et~al.}
  AFM-based high-throughput nanomechanical screening of single extracellular
  vesicles. \emph{Analytical chemistry} \textbf{2020}, \emph{92},
  10274--10282\relax
\mciteBstWouldAddEndPuncttrue
\mciteSetBstMidEndSepPunct{\mcitedefaultmidpunct}
{\mcitedefaultendpunct}{\mcitedefaultseppunct}\relax
\EndOfBibitem
\bibitem[Perissinotto \latin{et~al.}(2021)Perissinotto, Rondelli, Senigagliesi,
  Brocca, Alm{\'a}sy, Botty{\'a}n, Merkel, Amenitsch, Sartori, Pachler,
  \latin{et~al.} others]{perissinotto2021structural}
Perissinotto,~F.; Rondelli,~V.; Senigagliesi,~B.; Brocca,~P.; Alm{\'a}sy,~L.;
  Botty{\'a}n,~L.; Merkel,~D.~G.; Amenitsch,~H.; Sartori,~B.; Pachler,~K.,
  \latin{et~al.}  Structural insights into fusion mechanisms of small
  extracellular vesicles with model plasma membranes. \emph{Nanoscale}
  \textbf{2021}, \emph{13}, 5224--5233\relax
\mciteBstWouldAddEndPuncttrue
\mciteSetBstMidEndSepPunct{\mcitedefaultmidpunct}
{\mcitedefaultendpunct}{\mcitedefaultseppunct}\relax
\EndOfBibitem
\end{mcitethebibliography}
\end{document}


\section{Extracellular vesicles size characterisation}
\begin{figure}[H]
\includegraphics[width=0.9\textwidth]{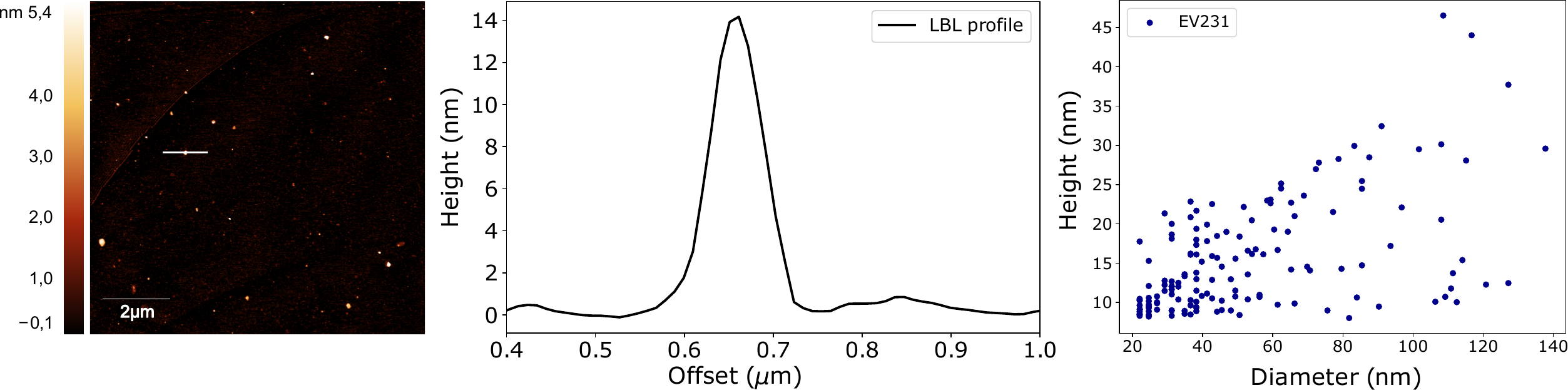}
\caption{AFM size characterisation of sEVs from MDA-MB-231 cell line, imaged in PBS 1X buffer over MICA substrate.}\label{fig:Figure1s}
    \end{figure}

For AFM analysis of individual sEVs isolated from MDA-MB-231 cell line, a freshly cleaved mica of $10\;mm$ in diameter with $0.15-0.21\;mm$ thickness (from Micro to Nano) was first incubated with $30\;\mu L$ of Poly-L-Ornithine (from Sigma-Aldrich) for $20\;min$; after an extensive washing with Milli-Q $H_2 O$, $30\;\mu L$ of sEVs were let to incubate for $20\;min$ and then gently rinsed with $50\;\mu L$ of $10\;\unit{mM}$ PBS before AFM imaging. \\
The AFM size characterisation is reported in Figure \ref{fig:Figure1s}S. From the size distribution reported in the scatter plot, sEV's typical height was $15.43\pm5.77\;nm$, with a mean diameter equal to $49.73\;\pm\;18.24\;nm$, calculated on a number of $112$ vesicles. These values resides in the typical range that can be found in literature \cite{perissinotto2020multi,ridolfi2020afm}.
\captionsetup{labelformat = suppl}
\section{MDA-MB-231 sEVs affinity test with single component SLB}
\begin{figure}[H]
\includegraphics[width=0.9\textwidth]{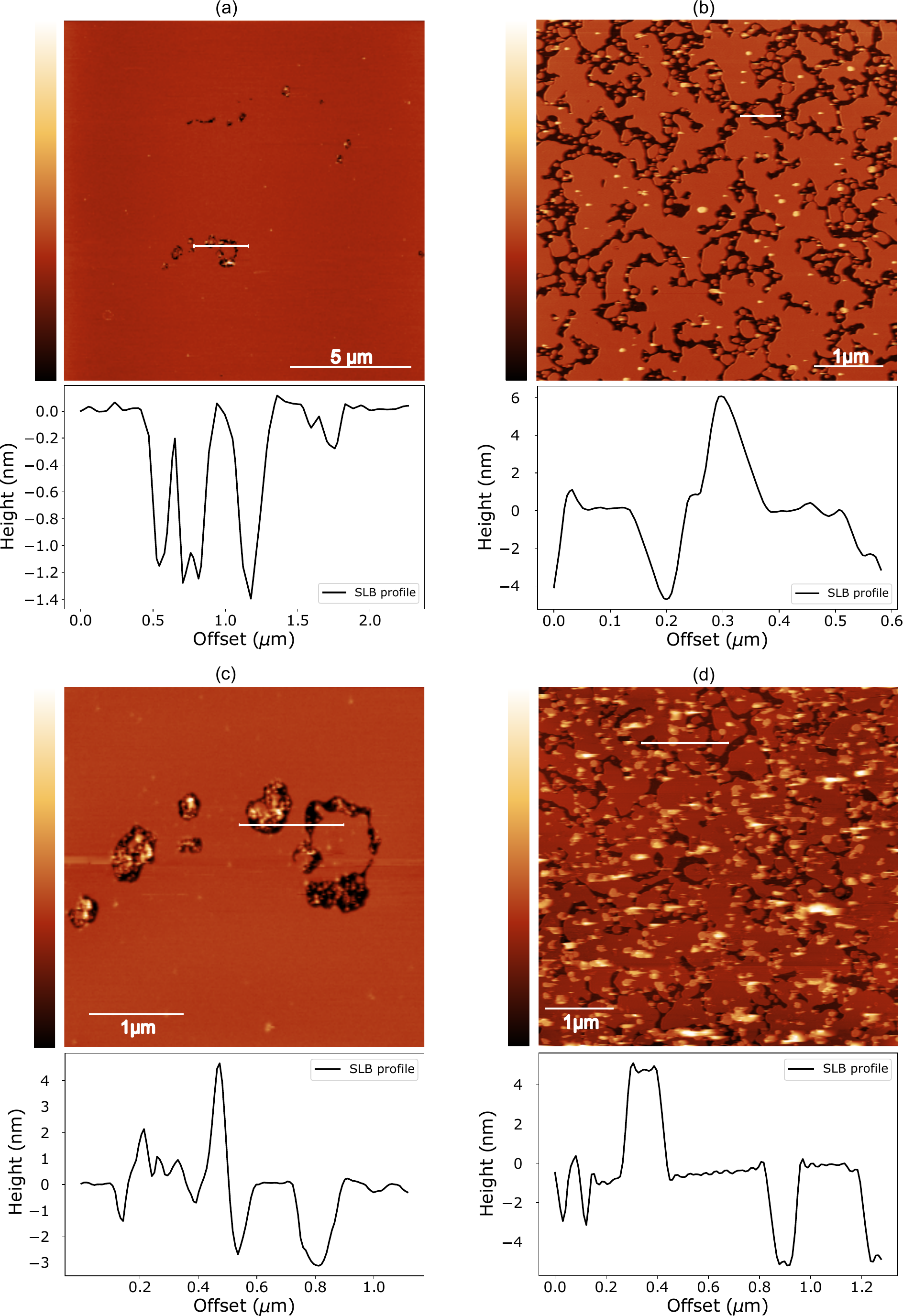}
\caption{AFM topographic images of DOPC and DPPC SLB before (a,b) and after (c,d) sEVs (MDA-MB-231 cell line) interaction with corresponding height profiles, acquired at $27^\circ C$ in Tris buffer $10\;mM$.}\label{fig:Figure2s}
    \end{figure}

To confirm the hypothesis of the sEV's (from MDA-MB-231 cell line) preferential mixing with the $S_o$ phase of the SLB, the vesicles have been tested in their interaction with a single component SLB composed of DOPC (Figure \ref{fig:Figure2s}Sa) or DPPC (Figure \ref{fig:Figure2s}Sb). In the first composition, as expected and reported in Figure \ref{fig:Figure2s}Sc, sEVs colocalize with the DOPC defects, without interacting with the surrounding SLB. The opposite behavior is reported in Figure \ref{fig:Figure2s}Sd for the DPPC SLB where the sEVs cover the whole area available. As previously reported, sEVs mixing with DPPC lead to the formation of an intermediate phase, regular in height, with a thickness comparable to the total thickness of the bilayer. 
\captionsetup{labelformat = suppl}
\section{UC-MSC sEVs interaction with model membrane}
\begin{figure}[H]
\includegraphics[width=0.9\textwidth]{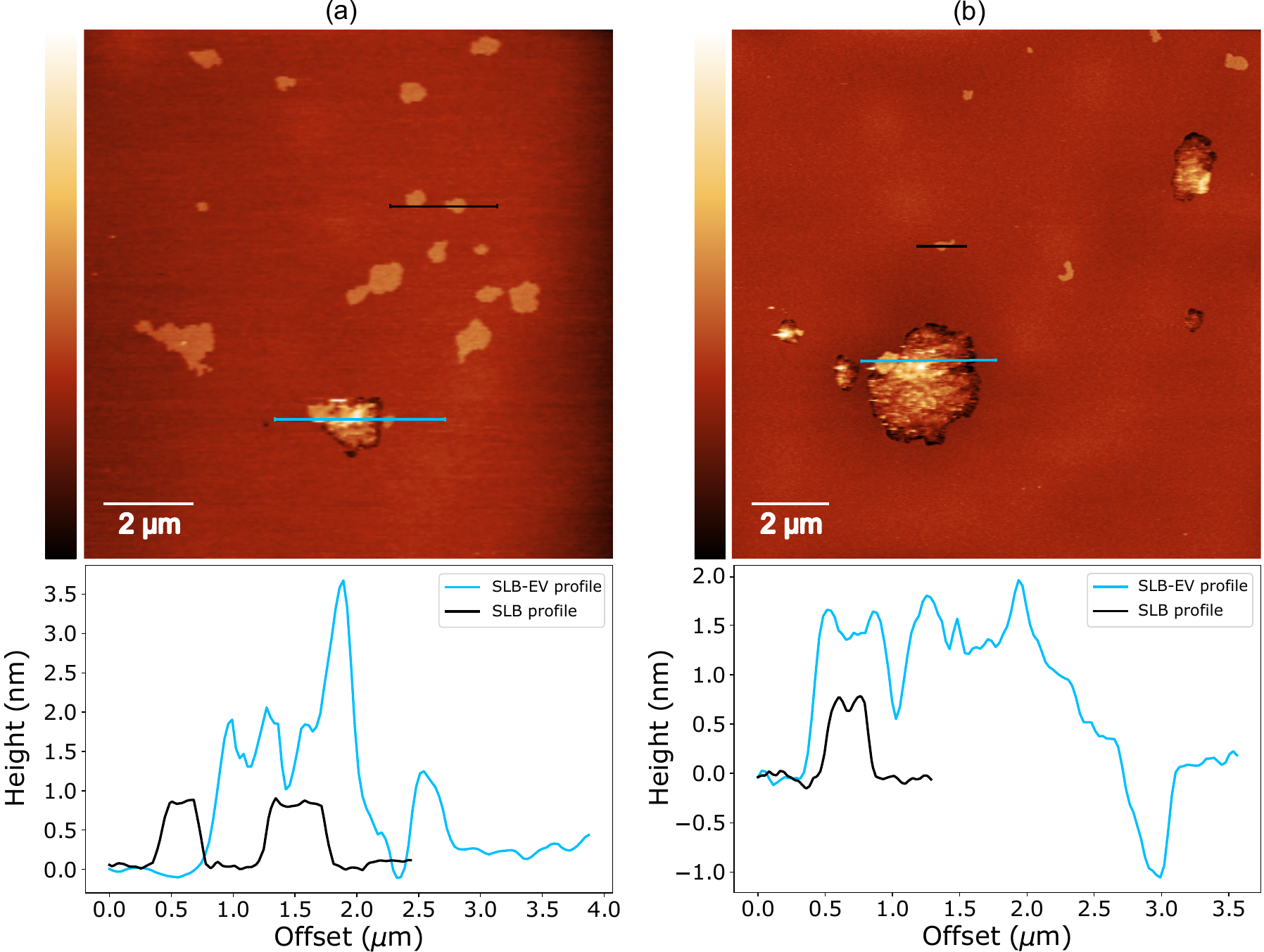}
\caption{AFM topographic images of the sEVs (UC-MSC cell line) interaction with SLB made of DOPC/SM $2:1$ (m/m) with $17\;mol \%$ Chol, and corresponding cross-sectional line profiles. Images are acquired at $27 ^\circ C$ in TRIS buffer $10\;mM$, with a time lapse of $40\;min$.}\label{fig:Figure3s}
    \end{figure}

To study the differences in the sEVs interaction process as a function of the sEV's origin, sEVs isolated from the human umbilical cord-derived mesenchymal stromal cells (UC-MSCs) have been added to the SLB composed of DOPC and SM with $17\;mol \%$ of Cholesterol. From the AFM image reported in Figure \ref{fig:Figure3s}Sa, it can be observed that the sEV's mixing is localised at the interphase of the $L_o$/$L_d$ domains, forming local invaginations with an average value of $0.5\;nm$ at the level of the $L_d$ phase. This result is in agreement with the work proposed by our group \cite{perissinotto2021structural} for the same sV sample, where it was attributed to the formation of a mixed phase due to sEV fusion with the SLB. A time-resolved analysis was also performed to track the evolution of the sEV mixing with the SLB. In Figure \ref{fig:Figure3s}Sb is reported the same area was analysed after $40\;min$ from the sEV uptake. It can be noticed a positive growth of the area occupied by sEVs patches that is accompanied by a dramatic decrease in the $L_o$ domains. However, the process evolution differs from what can be observed for the same SLB composition in interaction with the sEVs isolated from the MDA-MB-231 cell line, as reported in the main text. In the latter, the $L_o$ melting was not characterized by any significative increase over the time of the area occupied by the sEV's patches, while here this phenomenon is more pronounced. This can be attributed to intrinsic differences in the sEV's composition and mechanisms of interaction with the proposed model system.\\
\captionsetup{labelformat = suppl}
\bibliography{referencesSup}